\documentclass[a4paper, amsfonts, amssymb, amsmath, reprint, showkeys, nofootinbib,superscriptaddress,twoside, english,onecolumn, pra]{revtex4-2}

\usepackage[english]{babel}
\usepackage[utf8]{inputenc}
\usepackage[colorinlistoftodos, color=green!40, prependcaption]{todonotes}
\usepackage{lipsum}

\usepackage[pdftex, pdftitle={Article}, pdfauthor={Author}]{hyperref} 

\usepackage[normalem]{ulem}

\definecolor{FF}{RGB}{14, 159, 14}

\usepackage{graphicx}
\usepackage{subfigure}
\usepackage{tabulary}
\usepackage{amsmath}
\usepackage{color} 
\usepackage{float}
\usepackage{amsfonts}
\usepackage{amssymb}
\usepackage{mathtools}
\usepackage{dsfont}
\usepackage{eucal}
\usepackage{braket}
\usepackage[left=2cm,right=2cm,top=3cm,bottom=2.5cm]{geometry}
\usepackage{hyperref}
\hypersetup{
    colorlinks,
    linkcolor={black},
    citecolor={black},
    urlcolor={black}
}
\usepackage{caption}
\usepackage{multirow}
\usepackage{enumitem}
\usepackage{booktabs}
\setlist[itemize]{leftmargin=*}

\begin{document}
\title{Biophysical EPR Using Superconducting Resonators}

\author{Austin R. Gamble Jarvi}
\affiliation{High Q Technologies, Inc., Waterloo, Ontario, Canada}

\author{Hamid R. Mohebbi}
\affiliation{High Q Technologies, Inc., Waterloo, Ontario, Canada}
\affiliation{The Institute for Quantum Computing, University of Waterloo, Waterloo, Ontario, Canada}

\author{Ishit Raval}
\affiliation{High Q Technologies, Inc., Waterloo, Ontario, Canada}

\author{Omari Culzac}
\affiliation{High Q Technologies, Inc., Waterloo, Ontario, Canada}

\author{Jack Erickson}
\affiliation{High Q Technologies, Inc., Waterloo, Ontario, Canada}

\author{Sameh Hegazy}
\affiliation{High Q Technologies, Inc., Waterloo, Ontario, Canada}

\author{Don Carkner}
\affiliation{High Q Technologies, Inc., Waterloo, Ontario, Canada}

\author{Andrew Wiles}
\affiliation{High Q Technologies, Inc., Waterloo, Ontario, Canada}

\author{David G. Cory}
\affiliation{The Institute for Quantum Computing, University of Waterloo, Waterloo, Ontario, Canada}
\affiliation{Department of Chemistry, University of Waterloo, Waterloo, Ontario, Canada}

\author{Troy W. Borneman}
\email{troyborneman@gmail.com}
\affiliation{High Q Technologies, Inc., Waterloo, Ontario, Canada}
\affiliation{The Institute for Quantum Computing, University of Waterloo, Waterloo, Ontario, Canada}

\date{\today} 


\begin{abstract}
\noindent We present innovations that enable the use of superconducting resonators for high sensitivity, high bandwidth pulsed electron paramagnetic resonance (EPR) measurements on biologically relevant samples with enhanced stability and throughput. A custom-built X-band pulsed EPR spectrometer with AWG and digital IF capability generated by an FPGA was used to control a novel patterned thin film planar superconducting microstrip resonator capable of generating Rabi fields sufficient to achieve 6 ns $\pi$/2 Gaussian pulses using a 100 W solid-state HPA. The system allows automated sequential calibration, measurement, and analysis of five 3.5 $\mu$L samples contained in a sample cartridge. Performance was validated through measurements of double electron-electron resonance (DEER) distances in a variety of spin-labeled protein samples with biologically relevant concentrations, including measurements below 10 $\mu$M. The results enable broadening the scope of applications for both superconducting resonators and the use of EPR in biotechnology.
\end{abstract}

\maketitle

\section{Introduction}

Recent efforts to develop large-scale quantum computing systems have led to significant advances in quantum device technology. As a result, applications of these devices outside of information processing, communication, and sensing have become practical. For example, superconducting resonators have been used in electron paramagnetic resonance (EPR) measurements to demonstrate unprecedented spin sensitivity \cite{bienfait_reaching_2016,ranjan_electron_2020} on samples embedded in the device substrate or manually placed directly on top of the device. Biophysical EPR, in particular, is an important and extensive sub-field of EPR that imposes strict demands on sample placement and resonator design, including the requirement of loading pre-frozen samples from a room-temperature environment onto a pre-cooled device held at cryogenic temperatures ($< 77$ K).

Within the field of biophysical EPR, pulsed dipolar spectroscopy (PDS) has gained prominence as a powerful integrative structural biology tool. PDS is essential for studying dynamics and disorder in biomacromolecules \cite{jeschke_contribution_2018,wunnicke_synergetic_2017}, providing complementary information to other structural techniques, including Cryo-EM \cite{de_val_understanding_2023,stock_cryo-em_2018}, NMR \cite{fan_applications_2016}, X-ray crystallography, and FRET \cite{klose_resolving_2021}, and adding constraints to modeling strategies such as AlphaFold \cite{wu_modeling_2025,del_alamo_integrated_2022} and Molecular Dynamics simulations \cite{ettema_shared_2026,ma_swapped_2026} to produce more reliable structures for flexible and disordered regions. Further applications of PDS in biophysics include in-cell measurements \cite{bonucci_-cell_2020}, measurement of conformational dynamics of important binding sites \cite{reichenwallner_electron_2021}, and characterization of transcription processes in genetics \cite{hofmann_use_2022}. 

Superconducting resonators have found use in a variety of spin-based quantum technologies, including quantum information processing and sensing systems \cite{morton_storing_2018}, and traditional EPR specroscopy systems \cite{artzi_superconducting_2022,akhmetzyanov_electron_2023}. These applications take advantage of the high concentration-independent per-spin sensitivity of superconducting devices. However, for use in biophysical EPR, superconducting devices must satisfy a unique set of requirements: sufficient device bandwidth and power-handling capability to permit excitation and detection of broad linewidth spin-labels \cite{jeschke_deer_2012}; and relatively high sample volume to allow detection of samples at physiologically relevant concentrations ($< 10$ $\mu$M). In these requirements there is a contradiction: the inherent per-spin sensitivity advantage of superconducting resonators partially lies in their small mode volume (high filling-factor), generally requiring small sample volumes ($<1$ $\mu$L) that sacrifice concentration sensitivity relative to traditional EPR resonators (15 - 100 $\mu$L). Furthermore, the large drive currents necessary to excite broad linewidth samples generally exceed the critical current, $I_c$, of the device. 

The EPR system presented in this work was specifically designed for use in biophysical EPR, built around a novel superconducting microstrip resonator that enables the measurement of 3.5 $\mu$L vitrified samples with molecular concentration less than 10 $\mu$M. In addition to the benefit of material economy, the ability to measure small samples enables the use of multi-sample cartridges, improving measurement throughput and reducing bias in systematic studies that seek to isolate a single independent variable, including time-resolved studies \cite{schmidt_time-resolved_2022}. The stability of the system was optimized using a custom-built spectrometer with a digital intermediate frequency (IF) generated by a field-programmable gate array (FPGA) and operation of the resonator in a cryogen-free closed-cycle cryostat held under vacuum, stabilizing the resonance frequency. The FPGA was also used as an arbitrary waveform generator (AWG) and digitizer time-locked to the transmission system with 1 ns resolution and 500 MHz instantaneous analog bandwidth. Similar spectrometer architectures have recently been used in EPR \cite{kaufmann_dac-board_2013,shi_x-band_2018}. The operation of the resonator at cryogenic temperatures sets the noise floor well below room-temperature, allowing optimal utilization of cryogenic amplifiers \cite{kalendra_x-_2023,kalendra_q-band_2023,simenas_sensitivity_2021,twig_cryogenic_2012,pfenninger_noise_1995,narkowicz_cryogenic_2013}. Particular attention was also paid to automation and software design that enable a user unfamiliar with EPR to perform unattended sequential measurement and analysis of up to five samples. 

\begin{table}[hbt!]
\centering
\caption{System Specifications}
\begin{tabular}{lc}
\midrule
Operating Frequency & 9.5 GHz \\
Operating Field & up to 550 mT \\
Pulse Bandwidth & 500 MHz \\
Pulse Resolution & 1 ns \\
Delay Resolution & 1 ns \\
Digitizer Resolution & 1 ns \\
Conversion Efficiency & 5 G/$\sqrt{W}$ \\
Sample Temperature & 15 - 89 K \\
HPA Power & 100 W \\
Phase Stability & $\pm 1$ deg \\
Sample Size & 3.5 $\mu$L \\
\bottomrule
\end{tabular}
\label{tableIntro}
\end{table}

\section{Device Design}
The sensitivity of an EPR resonator is determined by the intensity of the detected voltage \cite{hoult_signal--noise_1976,rinard_absolute_1999},
\begin{equation}
    V= \chi''\eta Q\sqrt{P_AZ_0}
\end{equation}
where $\chi''$ is the RF susceptibility, $\eta$ is a filling factor that describes how well the sample volume is matched to the volume of the cavity mode, $Q$ is the resonator quality factor, $P_A$ is the applied microwave power, and the system is taken to be matched to $Z_0=50$ $\Omega$. A modern expression based on a bilinear photon exchange Hamiltonian operating jointly on a spin ensemble of size $N$ and cavity with strength $g=g_0 \sqrt{N}$ is often used in the literature due to the prevalence of new applications in cavity quantum electrodynamics (QED) measurements with spin ensembles \cite{albertinale_detecting_2021,ranjan_pulsed_2020,wang_single-electron_2023}. The single-spin coupling strength, $g_0$, is proportional to resonance frequency, $\omega_0$, inversely proportional to cavity volume, $V_c$, and may be held in correspondence with the standard filling factor expression: 
\begin{equation}
g_0 = \frac{\mu_B}{\hbar}\sqrt{\frac{2\mu_0\hbar\omega_0}{V_c}}
\end{equation}
Typical values for standard three-dimensional EPR resonators are $g_0\sim0.01$ Hz, while the superconducting resonator device used in this work provides $g_0=0.1-0.2$ Hz. Coupling strengths of several kHz have been reported in specific circumstances where the ensemble is placed directly on the surface of the resonator \cite{albanese_radiative_2020,ranjan_pulsed_2020}. In addition to increased sensitivity, a higher $g$ also provides better conversion efficiency, with the Rabi frequency being given as $\omega_1=2g|\mathcal{E}|/\kappa$ for a classical drive strength of $\mathcal{|E|}$ \cite{alsing_spontaneous_1991}. The sensitivity of a superconducting resonator EPR system, defined as the minimum number of detectable spins, has been shown to be equivalent to the classical expression \cite{bienfait_reaching_2016,blank_recent_2017} and is given by
\begin{equation}
    N_\text{min}^\text{Abs}=\frac{1}{gp}\sqrt{\frac{\kappa n}{T_E}},
\end{equation}
where $p=\tanh\left[{\frac{\hbar\omega}{k_B T}}\right]$ is the thermal polarization of the spin ensemble, $n$ is the number of noise photons, and $\kappa=\frac{\omega}{2Q}$ is the bandwidth of the resonator. For the sake of argument, we are neglecting the complex subspace structure associated with finite temperature spin ensembles interacting with a cavity, described by the Tavis-Cummings model \cite{tavis_exact_1968,gunderman_lamb_2021}, and simply treat $g$ as a proportionality constant operating in the Dicke subspace that dictates the conversion efficiency between microwave power and the resonator magnetic field.

The intrinsic per-spin sensitivity advantage of superconducting resonators lies in a few factors: a large internal quality factor, $Q_\text{int}$, limits noise bandwidth through small $\kappa$; the ability to pattern robust planar structures and place samples near the device surface leads to large values of $g$; and operation of the device at low-temperature provides a low fundamental noise floor, corresponding to small $n$. To obtain optimal concentration sensitivity,
\begin{equation}
S_{\text{conc}}=1/N_{\text{min}}^\text{conc} \propto g \sqrt{Q} V_\text{s}^{3/2},
\end{equation}
there is a trade-off between maximizing sample volume, $V_\text{s}$, to increase the number of spins contributing to the signal and minimizing cavity volume to obtain a large $g$. Maximizing concentration sensitivity requires engineering a cavity volume as large as possible while maintaining a strong spin-cavity coupling. 

\begin{figure*}[t!]
\centering
\includegraphics[width=11.4cm]{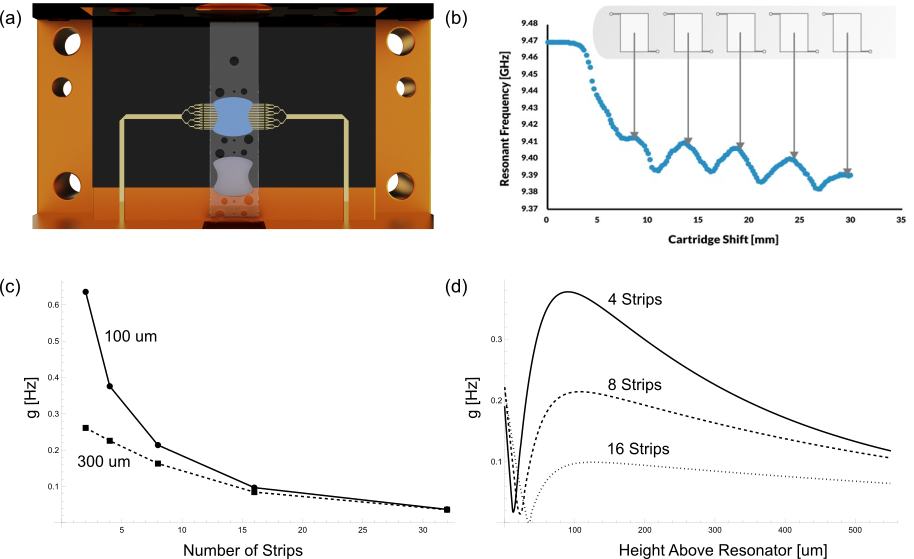}
\caption{Overview of superconducting resonator design. \textbf{(a)} Rendering of the 16-strip superconducting YBCO resonator used in this work with mode-matched sample cartridge placed in the homogeneous field region. The small size of the samples (3.5 $\mu$L) enables five sample cavities of roughly 3 mm x 2 mm x 400 $\mu$m each to be contained in a single cartridge. Adjustment of coupling into and out of the device is achieved through variation of a capacitive gap between the set of phased $\lambda/2$ resonators and a branching feedline that matches the impedance of each microstrip. The device provides a uniform microwave mode over a broad, relatively thin sample region. \textbf{(b)} Example resonance shift data from cartridge insertion with a schematic cartridge containing five sample cavities overlaid. The variable dielectric loading of the resonator by the borosilicate sample cartridge provides a means to locate specific sample cavities over the resonator. \textbf{(c)} Plot of the spin coupling strength in the center of the resonator at heights of 100 $\mu$m and 300 $\mu$m above the resonator surface versus the number of strips comprising the device. As expected, the field follows an inverse square-root relationship with the volume (surface area) of the resonator. \textbf{(d)} Stacked plot of the microwave field strength along the surface normal of the device for 4, 8, and 16 microstrips comprising the device. The volume of a sample contained within a region of equivalent homogeneity increases superlinearly with the number of microstrips.}
\label{fig:fig1}
\end{figure*}

To address these challenges, we used a novel superconducting microstrip resonator design \cite{mohebbi_composite_2014,benningshof_superconducting_2013}, comprising 16 phased $\lambda/2$ microstrip lines that produce a broad and relatively thin mode structure (Fig. \ref{fig:fig1}a). The volume of the resonator mode is effectively controlled by the number of strips comprising the device. In Fig. \ref{fig:fig1}b, a cross section of the coupling strength of the resonator to a single spin is shown at heights of 100 $\mu$m and 300 $\mu$m versus the number of strips. The volume of the cavity mode is expected to increase linearly with the number of strips, implying an expected behavior of $g(V_c)\propto 1/\sqrt{V_c}$. This is true near the resonator surface; however, the scaling of the coupling strength at increasing distances from the device no longer follows this behavior.

As shown in Fig. \ref{fig:fig1}c, the coupling strength as a function of height above the resonator surface depends strongly on the number of microstrips and, importantly, becomes more homogeneous over a thicker sample region. As a result, the sample volume, defined as the volumetric region where the field strength has a certain homogeneity, grows superlinearly with the number of microstrips. For example, the 16 strip resonator used in this work has a homogeneous sample volume, $V_s^{16}$, more than 16 times the sample volume for 1 strip, $V_s^1$, as summarized in Table \ref{table2}. The geometry of the sample was further optimized using HFSS simulations to yield a transverse bowtie profile with a thickness of 400 $\mu$m (Fig. \ref{fig:fig1}a), producing a sample volume of 3.5 $\mu$L. In addition to providing an advantage for volume-limited applications, the small sample volume, relative to common volumes of 20 - 100 $\mu$L, provides the opportunity to place multiple samples in a single sample cartridge (Fig. \ref{fig:fig1}). The cartridge used in this work contains up to five samples placed in cavities selectively etched into a laser-sensitized borosilicate substrate. The variable dielectric loading of the resonator as the cartridge is moved within the microwave mode volume creates a resonance shift that can be used to optimize the location of the sample cavity on the resonator (Fig. \ref{fig:fig1}b).   

The superconducting material of the resonator was chosen to enable measurements of vitrified samples at 50 K, the standard operating temperature of nitroxide free-radical moieties that are the most widely used spin-label in biophysical EPR. The common high-temperature superconductor material (HTS), YBCO, was chosen based on its critical temperature of 89 K and a history of use in EPR measurements \cite{bachar_nonlinear_2012,bonizzoni_microwave_2018,velluire_pellat_hybrid_2023}. An additional benefit of YBCO is a critical current large enough to handle drive powers necessary for fast pulsed control without kinetic inductance nonlinearity \cite{mohebbi_composite_2014,hincks_controlling_2015}.  The resonator is configured as a 2-port device operated in transmission mode, with coupling of microwave energy into and out of the device achieved through patterned capacitive gaps between the array of 16 microstrip resonators and a branching feedline structure that matches the impedance of each microstrip \cite{mohebbi_composite_2014}. To provide sufficient bandwidth to perform double resonance measurements, the device was overcoupled to a Q of roughly 80 by adjusting the capacitive feedline gap, providing a bandwidth of roughly 125 MHz while maintaining a high internal quality factor \cite{rinard_relative_1994}. Recent work has demonstrated that similar microstrip devices can be engineered to operate within a target parameter regime that exhibits tunable levels of non-Markovian backaction between the spin ensemble and the resonator, enabling the device used in this work to also find applications in cavity QED experiments \cite{gunderman_lamb_2021, gunderman_thermal_2025}.  

A final consideration in the design of the resonator was the operating frequency of the device and EPR system. The majority of recent PDS measurements are performed at Q-band frequencies (32-36 GHz) due to a significant increase in sensitivity relative to lower frequencies \cite{polyhach_high_2012}. The device presented in this work was designed for operation at X-band (9-10 GHz) to allow a larger inherent sample volume with a transmission line resonator, to reduce the impact of vortex-induced losses \cite{kwon_magnetic_2018}, and to enable excitation of the majority of the spin resonance spectrum. As a result of the lower operation frequency of our EPR system relative to the Q-band standard and the use of samples approximately ten times smaller, achieving a similar concentration sensitivity required the per-spin sensitivity of our X-band device to be approximately 120 times greater than a conventional Q-band resonator.

\begin{table}[hbt!]
\centering
\caption{Comparison of Resonator Geometries}
\begin{tabular}{lcccc}
Device & $\overline{g}$ & Sample Volume & $S_{abs}$ & $S_{conc}$ \\
\midrule
Microstrip-1 & 0.77 Hz & 0.024 $\mu$L & 6.38 & 0.0064 \\
Microstrip-4 & 0.44 Hz & 0.30 $\mu$L & 3.64 & 0.16 \\
Microstrip-8 & 0.25 Hz & 1.08 $\mu$L & 2.12 & 0.64 \\
Microstrip-16 & 0.12 Hz & 2.4 $\mu$L & 1 & 1 \\
\bottomrule
\end{tabular}
\\
Note: Sensitivities normalized to Microstrip-16. Sample volumes calculated assuming a rectangular sample cavity for ease of comparison.
\label{table2}
\end{table}

\section{Results}

\subsection*{Device Characterization}
\begin{figure*}[t!]
\centering
\includegraphics[width=11.4cm]{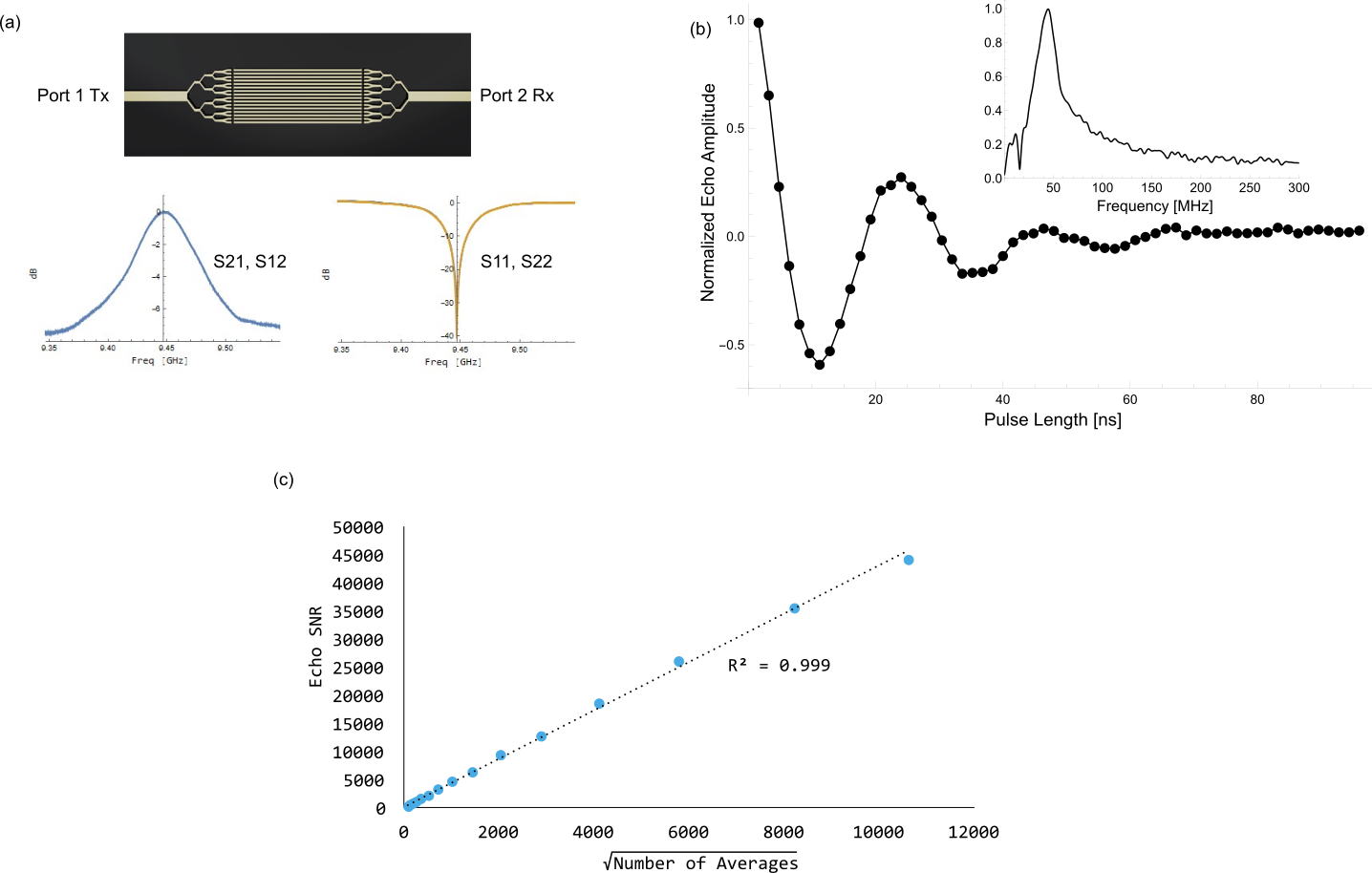}
\caption{Overview of system characterization. \textbf{(a)} Microscope photograph of the 2-port 16-strip 300 nm YBCO superconducting microstrip resonator structure used in this study with ports labeled. Each microstrip is 4.532 mm long and 70 $\mu$m wide with 70 $\mu$m separation between strips. The capacitive coupling gap is 60 $\mu$m for both ports. VNA S-parameter measurements show an X-band frequency of 9.447 GHz with an overcoupled Q of roughly 80, corresponding to a bandwidth of roughly 125 MHz. $S_{11}$ and $S_{22}$ data indicate the device is 50 $\Omega$ matched at both ports. \textbf{(b)} Nutation data with Fourier transform in the inset. Using approximately 4 W of power at the input to the resonator, 6 ns $\pi/2$ Gaussian pulses may be achieved with homogeneity matching HFSS simulations. \textbf{(c)} Plot of echo SNR versus square root of the number of averages demonstrating long-term stability of the system and the absence of systematic noise processes.}
\label{fig:fig2}
\end{figure*}

The results of a set of device characterization measurements are presented in Fig. \ref{fig:fig2}. Vector Network Analyzer (VNA) measurements demonstrate the high coupling efficiency of the device, with suppression of the reflected signal at resonance exceeding 40 dBc. Pulse nutation measurements indicate Rabi drive strengths that correspond to Gaussian $\pi$/2 and $\pi$ pulse lengths of 6 ns and 12 ns, respectively, enabling excitation of a significant portion of broad powder-like spectra associated with frozen spin-labeled biomolecules. 

\subsection*{System Stability}
The first widely-used EPR distance measurement sequence was a three-pulse double resonance (3p-DEER) scheme that detects the amplitude of a primary spin echo at an \emph{observer} frequency as a refocusing pulse is moved in time at a second \emph{pump} frequency \cite{milov_application_1981,milov_electron-electron_1984}. It was quickly recognized that short-time behavior of the dipolar modulation is unable to be detected in 3p-DEER due to spectrometer deadtime. Four-pulse double electron resonance (4p-DEER) \cite{martin_determination_1998,pannier_dead-time_2000} is the current standard sequence for biophysical EPR distance measurements. The sequence is robust to experimental error and resolves short-time dipolar modulation but lacks the sensitivity of more advanced methods due to the susceptibility of the observed refocused spin echo to $T_2$ signal loss \cite{bahrenberg_decay_2021}. 

Initial stability tests were performed by examining the scaling of a single echo SNR with respect to the number of averages. The theoretical scaling limit of $\sqrt{N}$ is often not realized in practice due to drift of the spectrometer phase, drift in the resonator frequency, or other instabilities that change the detected echo signal over time. As shown in Fig. \ref{fig:fig2}c, the ability of our device and spectrometer to coherently signal average is not practically limited.

Further stability tests during a long acquisition periods were performed by examining the phase behavior of the dipolar modulation signal in a 12 hour DEER measurement.
The results, shown in Fig. \ref{fig:fig4}b, demonstrate the absence of phase drift in the system and a phase stability of approximately $\pm$1 degree. We have not observed a significant deviation of the phase stability at any point for extended operation of the system without changing the temperature of the cryostat, suggesting that signal averaging over time periods of weeks or even months should be possible without loss of phase coherence. 

\subsection*{RELOAD DEER-Stitch with Echo Train Detection}
\label{sec:distancemeasurements}
The phase stability of our system enables us to take advantage of several important algorithmic adaptations that have been suggested for the standard DEER sequences: RELaxation-Optimized Acquisition Distribution (RELOAD), where the dipolar evolution time is changed over the course of acquiring a dipolar modulation trace \cite{milikisiyants_enhancing_2019}; DEER-Stitch, where two separate modulation traces corresponding to 3p and 4p are acquired and stitched together in post-processing \cite{lovett_deer-stitch_2012}; echo train detection that takes advantage of periodic refocusing of the detected echo through the application of a Carr-Purcell-Meiboom-Gill (CPMG) sequence \cite{mentink-vigier_increasing_2013,probst_inductive-detection_2017,ranjan_electron_2020,borneman_application_2010}; and the use of shaped pulses with designed bandwidths to maximize modulation depth while minimizing spectral overlap \cite{endeward_implementation_2023,lowe_optimizing_2024}. Utilizing these methods significantly increases SNR, but is not routinely done due to high implementation complexity and strict requirements on system stability.

\begin{figure*}[t!]
\centering
\includegraphics[width=11.4cm]{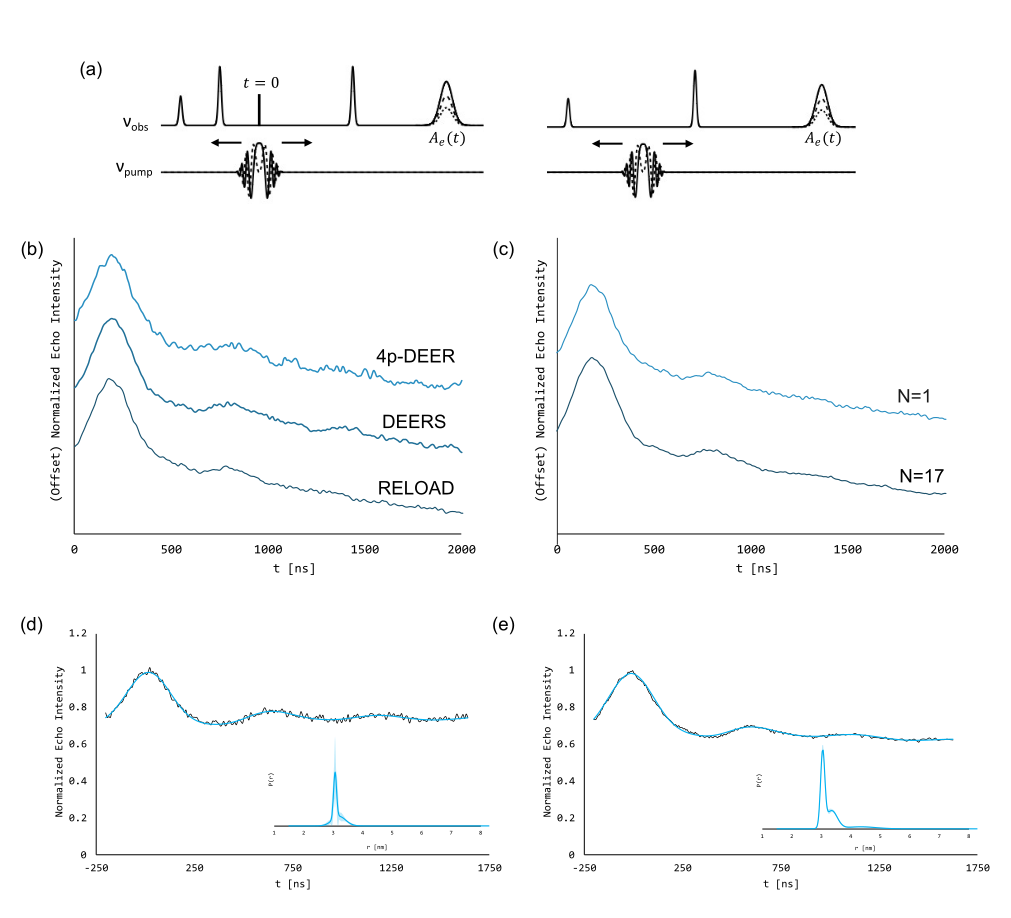}
\caption{Shaped pulse RDSE distance measurement data for T4-Lysozyme. \textbf{(a)} Pulse sequence schematics for 4p-DEER and 3p-DEER showing the usage of Gaussian observe pulses and adiabatic pump pulses. The dipolar evolution time is incremented commensurate with \textsuperscript{2}H-ESEEM (448 ns) according to the RELOAD algorithm. The 3p-DEER and 4p-DEER measurements are stitched according to the DEER-Stitch algorithm. The detected echo is refocused periodically by a CPMG sequence to increase SNR by averaging over a train of echoes in a single acquisition. \textbf{(b)} A comparison of standard shaped pulse 4p-DEER, DEER-Stitch, and DEER-Stitch+RELOAD (RDS) dipolar modulation traces for 90 $\mu$M T4L. The RDS methodology increases SNR by over a factor of 3. \textbf{(c)} A comparison of RDS and RDS+CPMG (RDSE) dipolar modulation traces of 90 $\mu$M T4L. The RDSE methodology increases SNR by over a factor of 6. \textbf{(d)} A dipolar modulation trace using RDSE for 5 $\mu$M T4L, demonstrating the capability to perform measurements at biologically relevant concentrations. \textbf{(e)} A dipolar modulation trace using RDSE for 45 $\mu$M T4L.}
\label{fig:fig3}
\end{figure*}

\begin{figure*}[t!]
\centering
\includegraphics[width=11.4cm]{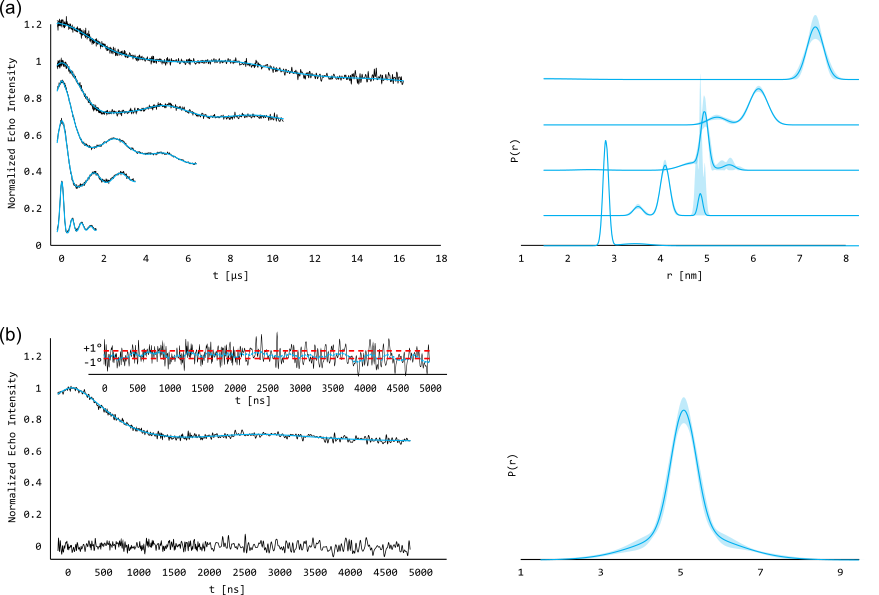}
\caption{Demonstrations of system measurement capabilities. \textbf{(a)} Measurement results of automated calibration, data collection, and distance analysis of 5 biradical ruler samples \cite{godt_how_2006} of varying distance in a single cartridge. The data was collected unattended after loading the cartridge and initiating the software. \textbf{(b)} Dipolar modulation data and processed distance distribution of a 25 $\mu$M YopO sample similar to that used in a round-robin benchmark study of Q-band EPR systems. Despite operating at X-band and with roughly x10 less sample volume, the SNR achieved by our system is comparable to values reported for Q-band in similar measurement times. The inset shows the phase of the quadrature detected signal over time, demonstrating signal phase stability of approximately one degree.}
\label{fig:fig4}
\end{figure*}

Distance measurements that utilize various advanced methodologies are shown in Fig. \ref{fig:fig3} for a spin-labeled T4-Lysozyme (T4L) sample \cite{balo_toward_2016,mchaourab_motion_1996,borbat_protein_2002}. The results are summarized in Table \ref{SNRtable}. The DEER sequence that uses shaped pulses, RELOAD, DEER-Stitch, and CPMG echo train detection is shown to provide a significant SNR benefit (x6) over a standard DEER sequence using shaped pulses. This advanced sequence, which we refer to as RDSE, was used for all distance measurements presented in this work and is implemented in our standard automated calibration, measurement, and analysis routines. RDSE was used to verify the ability of our system to measure sub-10 $\mu$M samples, an important baseline for biophysical EPR distance measurements. Fig. \ref{fig:fig3}d shows that the result of a RDSE distance measurement on a 5 $\mu$M T4L sample may be achieved in a relatively short period of time with sufficient SNR to provide a reliable distance analysis. 

\begin{table}[hbt!]
\centering
\caption{Performance Comparison of Advanced DEER Methodology}
\begin{tabular}{lccc}
Method & SNR \\
\midrule
1. Shaped Pulse 4p-DEER & 20 \\
2. DEER-Stitch & 57 \\
3. RELOAD DEER-Stitch (RDS) & 73 \\
4. RELOAD DEER-Stitch with Echo Train (RDSE) & 128 \\
\bottomrule
\end{tabular}
\\
SNR performance of various modified DEER sequences that enhance SNR. All sequences use Gaussian observe pulses \cite{teucher_improved_2018} and adiabatic pump pulses \cite{doll_adiabatic_2013}.
\label{SNRtable}
\end{table}

\subsection*{Automation}
\begin{figure*}[t!]
\centering
\includegraphics[width=11.4cm]{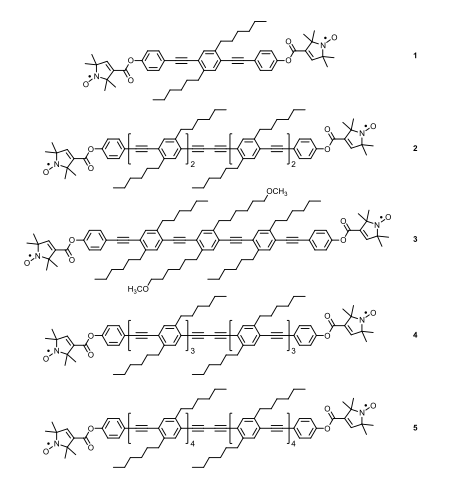}
\caption{Structures of dinitroxide rulers \cite{godt_how_2006} with distances listed in Table IV.}
\label{fig:fig5}
\end{figure*}

The high stability of our system and the dielectric similarity of each sample in our cartridges enable automated calibration routines, experiment setup, and data collection to run reliably without user input. Often, experiment setup and system tuning are the most time-consuming parts of a distance measurement, requiring careful calibration by a highly trained system operator. In Fig. \ref{fig:fig4}a and Table IV we present results obtained from automated calibration, measurement, and analysis of a sample cartridge containing five dinitroxide biradical ruler molecules of varying distance (Fig. \ref{fig:fig5}). The calibration time for each sample was roughly 15 minutes, with all measurements completed sequentially in less than 12 hours. The measurements presented, including distance analysis, were performed without user input at any point after entering a minimal set of sample information and initiating the cartridge measurement. The sample information includes a text description of the sample for the user, an expected range of distances for setting an optimal dipolar evolution time, and a target SNR or maximum measurement time. All relevant pulse sequence parameters are automatically determined by a calibration routine described in Materials and Methods.

\begin{table}[hbt!]
\centering
\caption{Biradical Sample Parameters}
\begin{tabular}{ccccc}
Sample & Distance & Concentration & SNR & Time \\
\midrule
\textbf{1} & 2.85 nm & 100 $\mu$M & 104 & 3.0 hr \\
\textbf{2} & 4.12 nm & 50 $\mu$M & 66 & 1.2 hr \\
\textbf{3} & 4.97 nm & 70 $\mu$M & 76 & 1.0 hr \\
\textbf{4} & 6.30 nm & 60 $\mu$M & 30 & 3.3 hr \\
\textbf{5} & 7.52 nm & 60 $\mu$M & 13 & 3.2 hr \\
\bottomrule
\end{tabular}
\\
Note: All measurements automated and performed sequentially
\label{table3}
\end{table}

\subsection*{Performance Benchmark}
To adhere to the performance standards of the biophysical EPR community, the presented system operating X-band system must achieve SNR and measurement time performance competitive with that of standard Q-band systems. A YopO sample similar to that used in a landmark round-robin study of the sensitivity and protocols of EPR \cite{schiemann_benchmark_2021} was used to benchmark system performance. In Fig. \ref{fig:fig4}b, we present the results of a distance measurement of a 3.5 $\mu$L sample of 25 $\mu$M YopO, achieving an SNR of 25 in a measurement time of 12.5 hours. This performance is in the neighborhood of the values reported in the round-robin study, indicating that our system does not significantly sacrifice performance to achieve enhanced automation, reliability, stability, and usability.

\section{Discussion}
We have demonstrated how an EPR system designed around a superconducting resonator device can be used to perform high stability, automated X-band biophysical EPR measurements with reduced sample volume. The concentration sensitivity achieved with this system is competitive with standard Q-band systems while using only 3.5 $\mu$L of sample. Achieving this performance required a factor of roughly 120 improvement in per-spin sensitivity. Of this, a factor of roughly 15 was provided intrinsically by the high filling factor and high internal Q of the superconducting device while the remaining factor of 8 was achieved through the utilization of advanced distance measurement methodologies, enabled by the overall stability of the system.

The use of a superconducting resonator allowed us to engineer an EPR system with long-term phase stability, enabling automated calibration, data collection, and analysis for unattended sequential measurement of up to five samples contained within a single cartridge. We are exploring several paths to further increase intrinsic sensitivity: moving to reflection mode measurements; moving to higher frequency and field; implementing pulse distortion compensation \cite{spindler_shaped_2012,doll_fourier-transform_2014}; and increasing resonator Q while using control sequences to account for the resulting distortions \cite{hincks_controlling_2015,borneman_bandwidth-limited_2012}. We are also implementing automated measurement and analysis routines for other EPR distance measurement techniques, including DQC \cite{mandato_double_2023}, RIDME \cite{milikisyants_pulsed_2009}, and SIFTER \cite{jeschke_dipolar_2000}, and other spin labels, including trityl radicals \cite{meyer_performance_2018} and copper \cite{gamble_jarvi_going_2021}.

\section{Materials and Methods}

\subsection*{Aqueous Sample Preparation}
9V1/131V1 T4 Lysozyme (T4L) samples were provided by the group of Oliver Ernst at the University of Toronto. T4L EPR samples were prepared to the desired concentration in a buffer of 100 mM NaCl, 50 mM Tris-HCl at pH 7.4 in D\textsubscript{2}O. 40\% v/v glycerol-d8 was added as a cryoprotectant.

588C/624C Yersinia outer protein O (residues 89-729) (YopO) was obtained from Genscript via custom protein synthesis and spin-labeled with (1-Oxyl-2,2,5,5-tetramethyl-3-pyrroline-3-methyl)methanethiosulfonate (MTSSL) and prepared per established protocols \cite{schiemann_benchmark_2021} with the following modifications: tris(2-carboxyethyl)phosphine hydrochloride (TCEP) was used in place of DTT, and Amicon Ultra – 0.5 mL Centrifugal Filters with 10-50 kDa cutoffs were used in place of desalting columns. Unless otherwise noted, all chemicals used were obtained from Sigma Aldrich.

EPR samples were loaded into sample cartridges via standard 1-10 µL pipette. Internal cavity volume is nominally 3.5 µL, however pipettes were set to load $\sim$4 µL of volume to compensate for filling channels and to ensure complete filling of the sample cavity. Full cartridges were flash frozen via submersion in liquid nitrogen before being inserted into the spectrometer.

\subsection*{Biradical Sample Preparation}
Nitroxide biradical rulers were prepared by Adelheid Godt and Miriam Hülsmann at the Max Planck Institute for Polymer Research and/or Bielefeld University. These biradicals were mixed with 10 times the amount of o-terphenyl-d14 (OTP) (Cambridge Isotope Laboratories) and the mixture was melted on a hot plate at 60\textsuperscript{o} C. The samples were loaded into sample cartridges using standard 1-10 µL pipettes as described above, with gentle heating applied to the loaded pipette tip by a hot air blower to maintain the sample in a liquid state. Once loaded into the cartridge, samples were gently heated by a hot air blower until liquid, then immediately flash frozen by submersion in liquid nitrogen. 

\subsection*{Sample Calibration Procedure}
Sample calibration and experiment setup are performed via a set of automated measurements that characterize the resonance frequency of the resonator, the sample spectrum, the sample $T_1$ and $T_2$ relaxation times, and a check of the single echo SNR.

\subsection*{RELOAD DEER-Stitch}
Several measurements are performed and concatenated in automated post-processing to compose a complete dipolar modulation trace. A first measurement is made using a 4p-DEER sequence to characterize the maximum amplitude and initial decay of the dipolar modulation. Next, a set of 3p-DEER measurements are performed with varying delay between the observe $\pi$/2 and $\pi$ pulses to mitigate relaxation. All delays are chosen to be commensurate with the period of the nuclear modulation in the sample. For example, for a nitroxide at X-band, deuterium nuclear modulation occurs at a frequency of 2.2 MHz, such that all delays should be integer multiples of 448 ns. As shown in Fig. \ref{fig:fig3}a, the initial 4p-DEER measurement will have delays of 448 ns and 896 ns, while the 3p-DEER measurements will have delays of 896 ns, 1344 ns, 1792 ns, etc. Each section of the dipolar trace is then stitched together. Extensive phase cycling is used (16-step for 4p-DEER and 4-step for 3p-DEER) to suppress unwanted echoes generated by the coherence of our observe and pump channels \cite{tait_coherent_2016}, which are defined digitally through the same hardware channel.   

\subsection*{Echo Train Detection}
For all measurements, the observed echo is refocused as many times as possible using a CPMG sequence, with the resulting echo train averaged together to increase SNR. Optimal operation of the CPMG sequence is highly sensitive to having finely tuned $\pi$ refocusing pulses that are phase aligned with the echo to be refocused \cite{borneman_application_2010,mandal_axis-matching_2013}. The length and phase of the refocusing pulses is fine-tuned in an automated calibration subroutine before each distance measurement.

\subsection*{Shaped Pulses}
\begin{figure*}[t!]
\centering
\includegraphics[width=11.4cm]{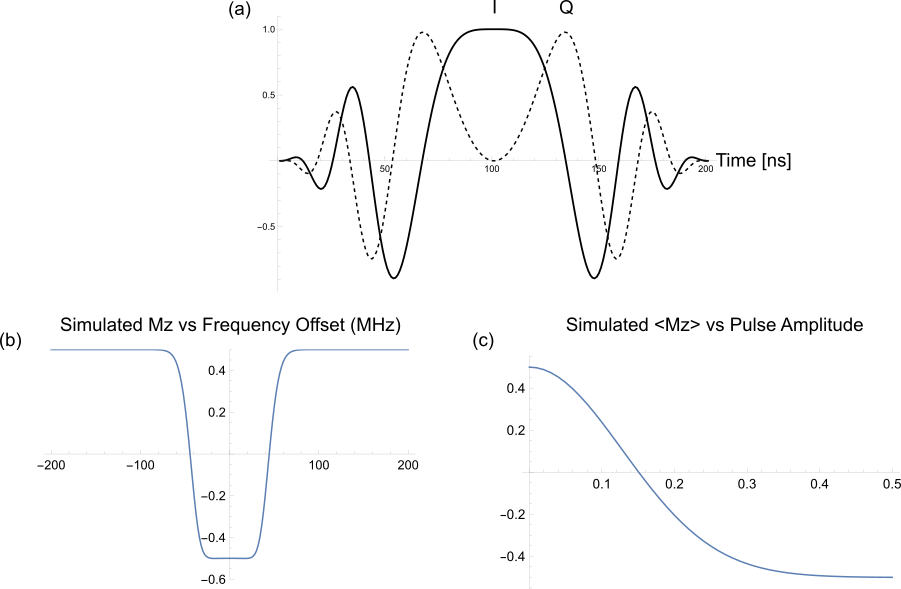}
\caption{Adiabatic WURST \cite{schops_broadband_2015} pulse waveform and simulated response for 200 ns pump pulse used in DEER measurements \textbf{(a)} IQ waveform. \textbf{(b)} Pulse response over resonance offsets. \textbf{(c)} Expectation value of Mz at resonance for varying pulse amplitude, indicating robustness to RF-inhomogeneity.}
\label{fig:adipul}
\end{figure*}

Shaped pulses are generated as a piecewise-constant arbitrary list of amplitudes and phases, $\{A_i,\phi_i\}$, where $i=-l_p/2$ to $l_p/2$ in 1 ns steps for an adiabatic pulse of length $l_p$ ns, and $i=-2l_p$ to $2l_p$ in 1 ns steps for a Gaussian pulse of full-width half-maximum length of $l_p$ ns. The functional form used for Gaussian envelope pulses is
\begin{align}
    A_i &= A_0 e^{-0.5(2 i/l_p)^2} \\
    \phi_i &= \phi_0
\end{align}
The functional form used for Adiabatic envelope pulses is
\begin{align}
    A_i &= A_0 \text{sech}(\beta i)(1-|\text{sin}(\pi i/l_p)|)^n\\
    \phi_i &= \text{BW} \space\text{log}(\text{cosh}(\beta i)/\beta)
\end{align}
where $\beta$ is a sweep rate in units of Hz, BW is a sweep width in units of Hz, and $n$ is a parameter that defines the curvature of the pulse amplitude. In Fig. \ref{fig:adipul} the IQ time-domain profile of a $l_p$ = 200 ns adiabatic pulse with $n=4$, BW = 1200 MHz, and $\beta=3.5$ MHz is shown \cite{schops_broadband_2015}, corresponding to the pump pulses used in the presented DEER data.

\section{Acknowledgements}
This work was partially supported by the Federal Economic Development Agency for Southern Ontario (FedDev) and the Strategic Innovation Fund (SIF) of Canada. We would like to thank Hassane Mchaourab (Vanderbilt), Richard Stein (Vanderbilt), Joerg Reichenwallner (University of Toronto), and Oliver Ernst (University of Toronto) for externally testing the EPR system and providing valuable feedback. We also would like to thank Adelheid Godt (Bielefeld University), Miriam Hülsmann (Bielefeld University), Jessica Besaw (University of Toronto), Joerg Reichenwallner, and Oliver Ernst for preparing and/or providing the samples used in this work.

\newpage

\section{References}

\bibliography{SCDEERrefs}

\newpage

\newpage

\end{document}